# The Illusion of the Perpetual Money Machine

*Didier Sornette and Peter Cauwels*
*ETH Zurich*
*Chair of Entrepreneurial Risk*
*Department of Management Technology and Economics*
*27 October 2012*

Eidgenössische Technische Hochschule Zürich
Swiss Federal Institute of Technology Zurich
**ETH**

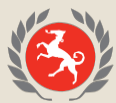

*Abstract:* We argue that the present crisis and stalling economy continuing since 2007 have clear origins, namely in the delusionary belief in the merits of policies based on a "perpetual money machine" type of thinking. Indeed, we document strong evidence that, since the early 1980s, consumption has been increasingly funded by smaller savings, booming financial profits, wealth extracted from house price appreciation and explosive debt. This is in stark contrast with the productivity-fueled growth that was seen in the 1950s and 1960s. This transition, starting in the early 1980s, was further supported by a climate of deregulation and a massive growth in financial derivatives designed to spread and diversify the risks globally. The result has been a succession of bubbles and crashes, including the worldwide stock market bubble and great crash of 19 October 1987, the savings and loans crisis of the 1980s, the burst in 1991 of the enormous Japanese real estate and stock market bubbles and its ensuing "lost decades", the emerging markets bubbles and crashes in 1994 and 1997, the LTCM crisis of 1998, the dotcom bubble bursting in 2000, the recent house price bubbles, the financialization bubble via special investment vehicles, speckled with acronyms like CDO, RMBS and CDS, the stock market bubble, the commodity and oil bubbles and the debt bubbles, all developing jointly and feeding on each other until the climax of 2008, which brought our financial system close to collapse. Rather than still hoping that real wealth will come out of money creation, an illusion also found in the current management of the on-going European sovereign and banking crises, we need fundamentally new ways of thinking. Governing is the art of planning and prediction. In uncertain times, it is essential, more than ever, to think in scenarios: what can happen in the future, and, what would be the effect on your wealth and capital? How can you protect yourself and your dearest against adverse scenarios? We thus end by examining the question "what can we do?" from the macro level, discussing the fundamental issue of incentives and of constructing and predicting scenarios as well as developing investment insights.

## Table of contents

# 1-A contemporary economic myth

*There is no use trying," said Alice. "One can't believe impossible things." "I daresay you haven't had much practice," said the Queen. "When I was your age, I always did it for half an hour a day. Why, sometimes I've believed as many as six impossible things before breakfast." ~Lewis Carroll*

Chasing fantasies is not the exclusive pastime of little girls in fairy tales. History is speckled with colorful stories of distinguished scientists and highly motivated inventors pursuing the holy grail of technology: the construction of a perpetual motion machine. These are stories of eccentric boys with flashy toys, dreaming of the fame and wealth that would reward the invention of the ultimate gizmo, a machine that can operate without depleting any power source, thereby solving forever our energy problems. In the mid-1800s, thermodynamics provided the formal basis on what common sense informs us: it is not possible to create energy out of nothing. It can be extracted from wood, gas, oil or even human work as was done for most of human history, but there are no inexhaustible sources.

What about wealth? Can it be created out of thin air? Surely, a central bank can print crispy banknotes and, by means of the modern electronic equivalent, easily add another zero to its balance sheet. But what is the deeper meaning of this money creation? Does it create real value? Common sense and Austrian economists in particular would argue that money creation outpacing real demand is a recipe for inflation. In this piece, we show that the question is much more subtle and interesting, especially for understanding the extraordinary developments since 2007. While it is true that, like energy, wealth cannot be created out of thin air, there is a fundamental difference: whereas the belief of some marginal scientists in a perpetual motion machine had essentially no impact, its financial equivalent has been the hidden cause behind the current economic impasse.

The Czech economist Tomáš Sedláček argues that, while we can understand old economic thinking from ancient myths, we can also learn a lot about contemporary myths from modern economic thinking. A case in point is the myth, developed in the last thirty years, of an eternal economic growth, based on financial innovations, rather than on real productivity gains strongly rooted in better management, improved design, and fueled by innovation and creativity. This has created an illusion that value can be extracted out of nothing; the mythical story of the perpetual money machine, dreamed up before breakfast.

To put things in perspective, we have to go back to the post-WWII era. It was characterized by 25 years of reconstruction and a third industrial revolution, which introduced computers, robots and the Internet. New infrastructure, innovation and technology led to a continuous increase in productivity. In that period, the financial sphere grew in balance with the real economy. In the 1970s, when the Bretton Woods system was terminated and the oil and inflation shocks hit the markets, business productivity stalled and economic growth became essentially dependent on consumption. Since the 1980s, consumption became increasingly funded by smaller savings, booming financial profits, wealth extracted from house prices appreciation and explosive debt. This was further supported by a climate of deregulation and a massive growth in financial derivatives designed to spread and diversify the risks globally. The result was a succession of bubbles and crashes: the worldwide stock market bubble and great crash of 19 October 1987, the savings and loans crisis of the 1980s, the burst in 1991 of the enormous Japanese real estate and stock market bubbles and its ensuing "lost decades", the emerging markets bubbles and crashes in 1994 and 1997, the LTCM crisis of 1998, the dotcom bubble bursting in 2000, the recent house price bubble, the financialization bubble via special invest-ment vehicles, speckled with acronyms like CDO, RMBS, CDS, …, the stock market bubble, the commodity and oil bubbles and the debt bubbles, all developing jointly and feeding on each other, until the climax of 2008, which brought our financial system close to collapse.

Each excess was felt to be "solved" by measures that in fact fueled following excesses; each crash was fought by an accommodative monetary policy, sowing the seeds for new bubbles and future crashes. Not only are crashes not any more mysterious, but the present crisis and stalling economy, also called the Great Recession, have clear origins, namely in the delusionary belief in the merits of policies based on a "perpetual money machine" type of thinking.

"The problems that we have created cannot be solved at the level of thinking we were at when we created them". This quote attributed to Albert Einstein resonates with the universally accepted solution of paradoxes encountered in the field of mathematical logic, when the framework has to be enlarged to get out of undecidable statements or fallacies. But, the policies implemented since 2008, with ultra-low interest rates, quantitative easing and other financial alchemical gesticulations, are essentially following the pattern of the last thirty years, namely the financialization of real problems plaguing the real economy. Rather than still hoping that real wealth will come out of money creation, an illusion also found in the current management of the on-going European sovereign and banking crises, we need fundamentally new ways of thinking. This will be the subject to the last part of this piece. But before, let us tell and document the magic story of the perpetual money machine of the last thirty years.

# 2-The turning point of the 1980s: from productivity to debt

## 2.1 GDP growth versus financial investment returns

Consider the following simple, almost naïve, question: is it sustainable for an economy, which expands at a real growth rate of 2−3 per cent per year, to provide a return of say 10−15 per cent per year averaged over all possible investment opportunities given to all investors?

This question specifically refers to what is called "the market portfolio" in the academic literature. This is a global, well-diversified portfolio that contains a representative basket of stocks, bonds, from sovereigns and corporates covering a wide range of risk exposures, commodities and real estate, directly and through exchange-traded-funds (ETF) and other financial instruments reproducing their cash flows, and even some private equity, hedge-fund shares and venture capital. In other words, it is the aggregation of the investments held by all the mutual funds, pension funds, and private investors, banks, sovereigns and so on. Can this portfolio really return more than the growth of the GDP?

The standard valuation models can help us to frame this fundamental question. Take for example the famous Gordon-Shapiro equation. This is based on the simple fact that corporate profits are either distributed to investors as dividends or are saved as retained earnings. There is a fundamental link between the return of investors, either in the form of a dividend yield or a capital increase, and the generation of profit of a company. When one applies this simple rule to the global economy, this means that the aggregate return of all financial investments, the return on the market portfolio, cannot grow faster than the real economy. In the end, behind every increased return on investment, there should be a creation of fundamental value.

Sure, there can be some companies or sectors with excellent potential because of new innovations, technology or the access to a new market. These can show a transient accelerated growth. Additionally, profits can also come from outside the country, from offshore investments and as such be disconnected from the performance of the internal economy. U.S. companies, operating internationally, have indeed derived, in the last decades, a larger and larger share of their profits worldwide outside the U.S. itself. But this effect accounts for at most a few percent of the huge difference between 2−3 and 10−15 percent. But overall, global wealth cannot grow faster than GDP in a sustainable way. In fact, any difference can only be explained through the existence of bubbles, which in this context can be understood as the transitory accelerating financial growth of a sector or a company that is not translated into real productivity gains.

However, in the last decades, it is quite clear that the balance has been violated repeatedly by the extraordinary expansion of the financial sphere. This phenomenon is illustrated in figure 1, which compares the real U.S. stock market appreciation as measured by the S&P500 stock index in the period 1952–2012 to the U.S. GDP, and figure 2, which shows the total U.S. household net worth compared with the U.S. stock market. The use of GDP as a universal yardstick is actually not trivial, given the many existing dynamical feedback loops, including the nature of the expectations of consumers and investors in the future, which are extremely difficult to disentangle. Nevertheless, we will use it here as a benchmark of real growth because, since WWII, it has grown, in real terms, in a narrow corridor around 3% per year. It behaves in such a regular and stable way that it makes sense as a standard to compare all other economic variables against. This is like the "standard candles" that are used in Astronomy to estimate the expansion of the universe.

The first observation is that GDP and U.S. stock market valuation have grown roughly at the same rate, supporting the above argument of their joint balanced growth on the long term. However, it is striking how the U.S. stock markets have exhibited three large periods of excessive valuations followed by periods of consolidation. The first period was associated with the so-called "tronic boom" of the 1960s. Closer to present, the fast acceleration of equity valuation since the mid-1990s was a clear sign of a massive financial market bubble, as shown in figure 1. After a quick deflation from 2000 to 2002, another stock market excess developed that ended in 2008. Figure 2 confirms this observation by presenting the evolution of the total household net worth in the U.S. expressed as a fraction of the GDP from 1952 to March 2012. This ratio was relatively stable, hovering between 300% and 350%, for the first 40 years in the graph. Since 1995, however, two major peaks towering above 450% can be observed. The first peak coincides with the peak of the dotcom bubble in 2000 that was followed by more than two years of strong bearish stock markets. The second even more impressive peak coincides with the top of the housing and stock market bubbles in 2007 followed by a correction of over 100% in units of GDP. The figure further compares the household wealth with the stock market. Note the strong co-evolution of the two curves from the mid-1980s onwards. This demonstrates that the total household net worth has been increasingly linked to financial market performance. Household wealth has basically been slaved to the bubbles and crashes that controlled the financial profits obtained from stocks market investments and real estate. Interestingly, the present wealth level at about four times GDP is still above the average level of the pre-1990 period. One can raise the question whether this is a sign of permanent improvements or whether more deflation is still to come.

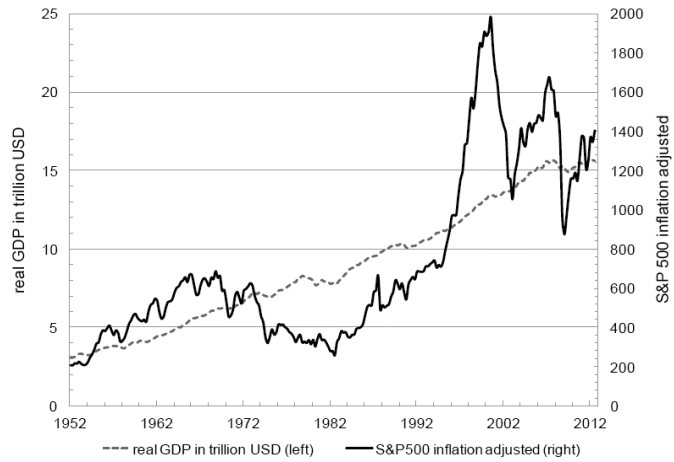

*Figure 1: Real GDP in 2012 U.S. dollars (dashed line) compared with real stock market value (continuous line) proxied by the S&P500 stock index from 1952 to 2012. The left and right vertical scales respect ratios so that the growth rates of GDP and S&P500 can be compared visually, source of data: Bloomberg*

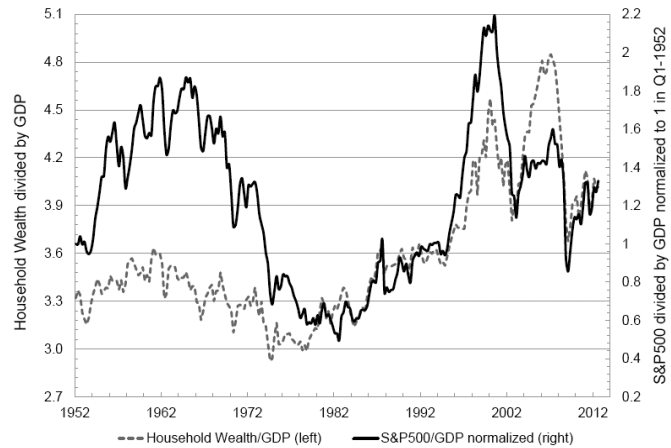

*Figure 2: The U.S. stock market performance (continuous line) and the household net worth in the U.S. (dashed line) in units of GDP from 1952 to Q1 2012. The household net worth includes real estate and financial assets (stocks, bonds, pension reserves, deposits, etc) net of liabilities (mostly mortgages). The graph uses the Fed's Q1 2012 Flow of Funds data (http://www.federalreserve.gov/releases/z1/current/default.htm).*

## 2.2 From productivity to consumption and debt

Figures 3 and 4 digested together offer an eye-opening piece of information complementing figure 1. Figure 3 shows the historical evolution of private consumption and wages as a percentage of GDP, aggregated over the U.S., the European Union and Japan since 1960. Until the mid-1980s, wages mainly funded consumption. Thereafter, consumption has outstripped wages and the gap has been increasing dramatically. This begs the following fundamental question: if not by wages, how has this increase in consumption been financed?

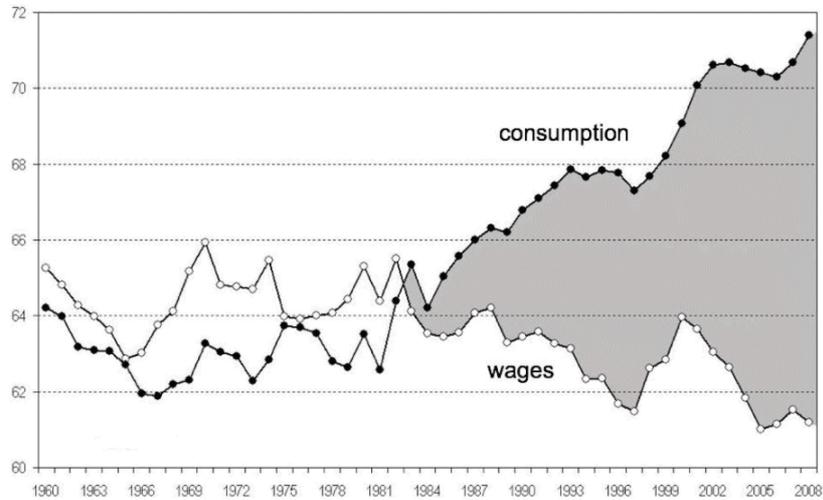

*Figure 3: The share of wages and of private consumption as a percentage of Gross Domestic Product (GDP) for the U.S., the European Union and Japan. Source of data and graphics: Michel Husson (http://hussonet.free.fr/toxicap.xls).*

Figure 4 gives a first explanation. It shows that households in the U.S., the European Union and Japan have increased their overall level of consumption by somewhat decreasing savings and mainly by extracting wealth from financial profits. Specific numbers for the U.S. alone confirm and even amplify this conclusion.

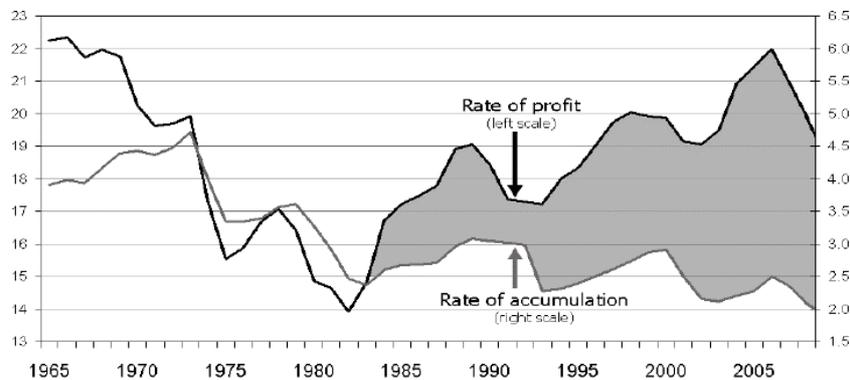

*Figure 4: The rate of profit in percent (left scale) and the rate of accumulation or savings in percent (right scale) for the U.S., the European Union and Japan. The rate of accumulation is defined as the growth rate of net capital. Source: Ameco Database, European Commission, http://tinyurl.com/ameco8. Reproduced from Michel Husson (http://hussonet.free.fr/toxicap.xls)*

The final piece of the puzzle is given in figure 5, which shows the evolution of the U.S. debt in percentage of GDP, defined as the sum of the Federal government debt plus private sector debt (households + firms), excluding social liabilities. One can observe its explosion since the mid-1980s. By 2003, it reached levels only seen 70 years earlier, in 1933, at the depth of the Great Depression. We will analyze this graph in more detail using our bubble models in section 3.2, and the result of this analysis can be found summarized in figure 16.

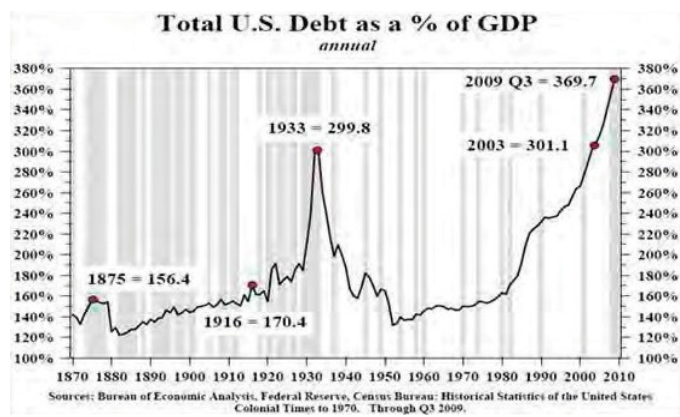

*Figure 5: The U.S. total credit market debt as a % of GDP (includes Government and Government Sponsored Entities, Households, Corporates and Financials). Source Hoisington Investment Management, Bureau of Economic Analysis, Federal Reserve, Census Bureau: Historical Statistics of the United States Colonial Times to 1970.*

All these pieces of data converge to the same crucial point: the mid-1980s have witnessed a fundamental change in regime, a clear break in the dynamics of our economic system. This coincides with what Alan Taylor, of the University of Virginia, calls the shift from the "Age of Money" to the "Age of Credit"[1]. A change he characterizes by an explosion of banks' balance sheets and a "mysterious" break in the fundamental macro-economic relationship between the growth of money and the growth of the real economy. In the previous paragraphs, we have described this regime shift as an increase in consumption only weakly offset by a decrease in savings, where the remaining balance is paid by debt and by financial profits either from stock market investments or so-called mortgage wealth extraction from house price appreciation.

The big question remains whether these financial profits were somehow translated into real productivity gains and, therefore, whether they were sustainable. As long as the incomes from financial assets are re-invested and kept in the financial sphere independently of the "real" economy, their prices can increase independently of any economic reality. But, in essence, financial assets represent the right to a share of some future surplus value, profit or revenue. Provided this right is not exercised, asset prices can continue their bubbly trajectory. However, as soon as it is exercised, it becomes subject to the law of value. At that moment, prices are judged against an expected fundamental value and suddenly it is remembered that one cannot distribute more real wealth than is produced.

The discrepancy between the exuberant inflation of the financial sphere and the more moderate growth of the real economy is the crux of the problem we are currently immersed in. This is further illustrated in the following two exhibits. Figure 6 presents the evolution with time of the Total Factor Business Productivity and shows that, since the early 1980s, productivity growth has been slowing down significantly. Figure 7 gives the relative wages of the U.S. financial industry compared to a benchmark index. If anything has increased since the beginning of the eighties in addition to consumption and debt, it is the income of financial professionals that has paralleled the growth of the financial sphere. Comparing figures 5 and 7, it is also striking to see how the wages in the financial industry follow the total debt level, an indication of the joint development of debt, financial investments and financial compensations.

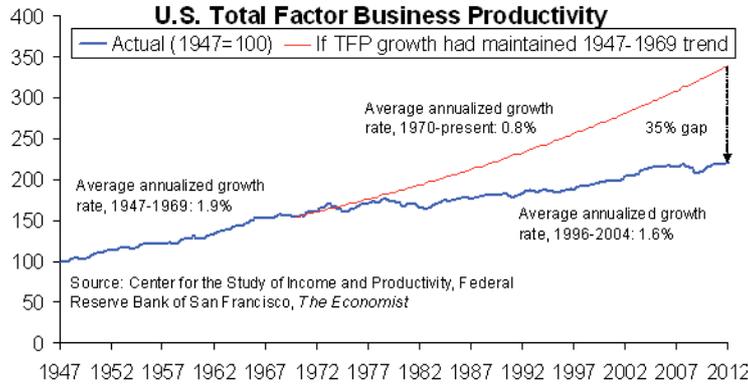

*Figure 6: The U.S. total factor business productivity since 1947. Source: Center for the Study of Income and Productivity; Federal Reserve Bank of San Francisco, the Economist (http://www.economist.com/blogs/freeexchange/2012/09/productivity-and-growth)*

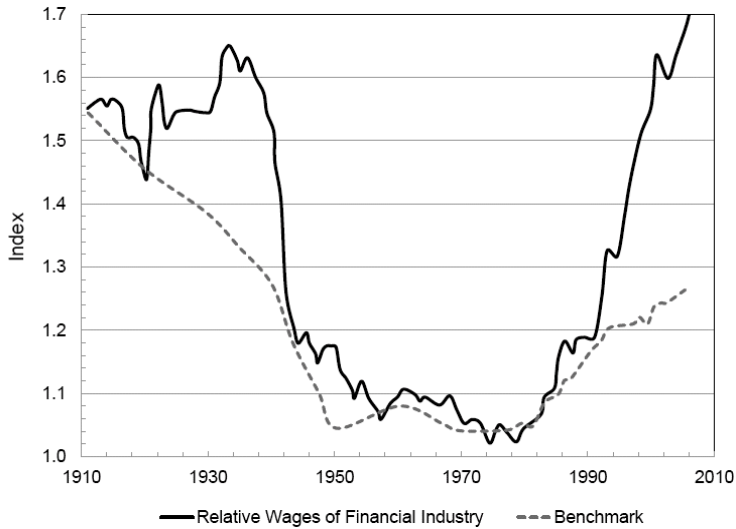

*Figure 7: Actual Relative Wages in the Financial Industry compared to a benchmark. Source: National Bureau of Economic Research, Wages and Human Capital in the U.S. Financial Industry: 1909 – 2006, Working Paper 14644 (http://www.nber.org/papers/w14644)*

The lack of recognition of the fundamental cause of the current crisis is symptomatic of the spirit of our time. Apparently, it is hard to accept that there is something more than just a downturn phase of a business cycle; that previous gains were not real, but artificially inflated values that have bubbled in the financial sphere, without anchor or justification in the real economy. In the last decade, banks, insurance companies, Wall Street as well as Main Street and many of us have lured ourselves into believing that we were richer. But this wealth was just the result of a series of self-fulfilling bubbles. The delusional spirit of the time is no better captured than by figure 8, which reproduces a study by the McKinsey Global Institute showing the change with time of the U.S. household debt as a percentage of gross disposable income. What is remarkable is the discussion found in the McKinsey article and illustrated in the figure that, as of the last quarter of 2011, households have deleveraged about half-way since the peak of 2008 towards the "normal", represented by the linear upward trending grey line. It is therefore implied that things are improving and, with a little more efforts, we will be back to "normal". But is a steady growth of relative debt really "normal"? In fact, for the twenty years from the mid-1960s to the mid-1980s, the relative debt ratio hovered around the 60 percent level, which is perhaps by coincidence the ratio of debt-to-GDP enshrined in the Maastricht treaty for the countries of the Eurozone. Thereafter, this ratio started to growth and accelerated until it peaked in 2008. A "normal" debt ratio of 100 percent in 2013 growing along the trend line implies a "normal" debt ratio of 140 percent in 2050 and of 180 percent in 2090! This thinking is really the most astonishing aspect of the recent developments. It is the signature of a fundamental transition to a new regime that we will analyze and project forward later.

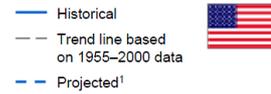

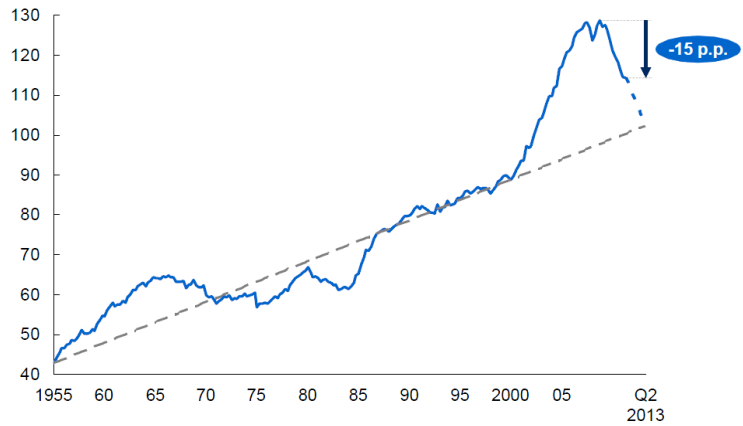

*Figure 8: U.S. household debt as percentage of gross disposable income. Reproduced from McKinsey Quarterly, publication of the McKinsey Global Institute, January 2012*[2]

But first, to obtain a deeper understanding of the current crisis, let us analyze how the last 30 years developed.

# 3-The roots of the crisis: three decades of perpetual money machine (1980–2007)

In the following, we will first go through the most common or often cited explanations, for the current crisis. Then, we will give our view, which is based on the contemporary myth of the perpetual money machine and the sequence of bubbles that it caused. This section may be a bit technical. Readers, who want to avoid this, can easily jump to section 4, where a summary of the main message of the quantitative and solidly documented present section is offered.

## 3.1 Common explanations of the financial crisis

Many explanations have been given, in the media and in the academic literature, for the current crisis. From a broad historical perspective, Harvard University historian Niall Ferguson argues that the financial crisis is an accelerator of an already well-established trend of relative Western decline. Others, like Raghuram Rajan, a finance professor at the Chicago Booth School of Business and former chief economist of the IMF, proposes that inequality has been the driving force behind the political decision to expand lending to (especially low-income) households, therefore seeding the subprime crisis. The same rhetoric can be developed for the sovereign European crisis rooted in the massive amount of lending that banks, encouraged by the European Central Bank, have extended to less developed European states since 1999. Mark Stein, from the School of Management at the University of Leicester, considers the crisis from a psychoanalytical vantage point. He suggests that the prevalent manic culture of denial, omnipotence, triumphalism and over-activity that characterizes the West created the conditions for the problems that led to the current crisis.

A more familiar explanation of the crisis is the expansion of the sales dominated industry, the "originate to distribute" culture, with its strong incentives to sell structured financial products. There has always been a major focus on innovation in the financial industry. However, with no one, at the top management level, really willing, or competent enough, to step on the brakes, and without proper balance by risk management, it can be said that the financialization bubble has been primarily pushed by the sales force. It is important to stress here, however, that this has been, to a large extent, a problem of incentives, as these sales were doing a good job within their own incentive structure. The growth of the hedge fund industry further increased the pressure on the traditional money managers to come up with similar rate of returns.

The corresponding financial innovation was embodied in particular in the extensive securitization of bad quality loans transformed in highly rated bonds, akin to transmuting lead into gold. These Asset Backed Securities were bought by institutional investors, banks, insurance companies and even money market funds. A whole new industry, called the shadow banking system (allowing off-balance sheet accounting), thrived on these toxic assets. As long as home prices continued to rise and interest rates stayed low, this shaky castle of cards somehow stayed in balance because the subprime borrowers had low interest payments and, in case they did default, their loan was backed by a property that had actually gained in value. This Ponzi scheme finally collapsed after the real estate bubble in the U.S. burst.

Financial innovation was further fueled by risk management models that showed how risks could be diversified away and made more liquid, in particular via the creation of derivatives. Advances in IT enabled to price everything, which gave an illusion of control and led to less common sense thinking about replication, hedging and liquidity risks. Besides, most of these models also ignored or under-estimated the linkage between different risks, by not making use of the new scientific insights, gained in other research fields in which it is well-known, that more diversification, increased homogeneity and stronger coupling leads to reduced mild risks (the diversification benefit) at the cost of drastically increased extreme risks (the curse of integration leading to system-wide risks).

Financial innovation and the shadow banking system were facilitated by a general consensus that less regulation and government overview was better for the economy. In line with the free market ideology advocated by Milton Friedman and Friedrich Hayek, a wave of financial deregulation set off in the 1980s, crowned in 1999 with the Gramm-Leach-Bliley act that repealed the Glass-Steagall act of 1933. As a result, the barriers between retail banks, commercial banks, investment banks and insurance companies were taken down. This increased the interconnectedness and therefore the susceptibility of the financial system to major catastrophic events. The deterioration of the quality of bank capital, by the substitution of common equity with creative hybrid capital, a blind trust in rating agencies, regulatory arbitrage, which is the polite name for using innovations to circumvent limits imposed by existing regulations, and a total absence of regulation of the shadow banking system further impaired the system's resilience.

## 3.2 Bubbles galore

Though many of these accounts touch on very deep insights and enlighten some fascinating aspects of the crisis, they merely describe symptoms as well as facilitating processes. In our view, the real fundamental mechanisms are associated with the change of regime that occurred approximately 30 years ago with the transition from a growth based on productivity gains to one based on debt explosion and financial gains. This is a symptom of what we previously referred to as the illusion of the perpetual money machine. By now, from the evidence provided by figures 1–8, should it not be striking that such an obvious point of discussion has not yet found its rightful place in the debate? Vaclav Havel, the Czech poet, dissident and politician, once wrote that education is the ability to perceive the hidden connections between phenomena. Let us take his advice at heart and examine more closely the big economic and financial trends of the past three decades to figure out the hidden connections that bind them together.

### 3.2.1 Universal bubble scenario and our original bubble model

A financial bubble is a curious beast: its meaning is accepted as obvious by the general public, yet its very existence is loudly debated in angry terms among experts. Arguably, almost any given adult met on the street would know exactly what one is and could cite examples in recent and distant history. The dotcom bubble ending in 2000 and the housing bubble recently ended would most likely be the most common examples given. More well-read people could cite the Dutch tulip mania in the 1600s and the South Sea Company of the 1700s. After that, the examples are less well-known but not because of their scarcity but just because most people are not interested in financial history and debates. This is changing.

Due to the recent global financial crisis, interest in bubbles among experts and non-experts alike is soaring. For the latter, a good bellwether is the frequency of the use of the word "bubble" in google searches and online news tracked by google[3]. Late 2007, at the time when the former U.S. Federal Reserve Chairman Alan Greenspan stated that "we've had a bubble in housing", the google search volume index for "bubble" began to climb dramatically (we do not imply causation here, merely correlation). This self-referential climb locally peaked before the end of the year 2007 but has remained consistently higher during the last five years. Clearly this is anecdotal evidence, but it is also very illuminating because this type of word-of-mouth spread of popularity is exactly one of the amplification mechanisms behind the growth of bubbles on which our methodology is based. And it makes complete sense: as more people actively seek out information about bubbles, more doubts are planted in their minds that they might be living in one and they think, "it is time to get out."

The largest housing boom of all time began its equally large bust at the end of 2007, the point at which the interest in the word "bubble" peaked (again we do not imply causation here, merely correlation).

The term "bubble" refers to a situation in which excessive future expectations cause prices to rise. For instance, during a house-price bubble, buyers think that a home that they would normally consider too expensive is now an acceptable purchase because they will be compensated by significant further price increases. They will not need to save as much as they otherwise might, because they expect the increased value of their home to do the saving for them. First-time homebuyers may also worry during a bubble that if they do not buy now, they will not be able to afford a home later. Furthermore, the expectation of large price increases may have a strong impact on demand if people think that home prices are very unlikely to fall, and certainly not likely to fall for long, so that there is little perceived risk associated with an investment in a home. What is the origin of bubbles? In a nutshell, speculative bubbles are caused by 1) precipitating factors that change public opinion about markets or that have an immediate impact on demand and 2) amplification mechanisms that take the form of price-to-price positive feedback: the larger the price, the higher the demand and ... the larger the price! This can be described by a universal scenario[4,5]. First, a novel opportunity arises. This can be a ground-breaking technology or the access to a new market. An initial strong demand from first-mover smart money leads to a first price appreciation. This often goes together with an expansion of credit, which further pushes prices up. Attracted by the prospect of higher returns, less sophisticated investors then enter the market. At that point, the demand goes up as the price increases, and the price goes up as the demand increases. This is the hallmark of a positive feedback mechanism. The behavior of the market no longer reflects any real underlying value and a bubble is born. The price increases faster and faster in a vicious circle with spells of short-lived panics until, at some point, investors start realizing that the process is not sustainable and the market collapses in a synchronization of sale orders. The crash occurs because the market has entered an unstable phase. Like a ruler held up vertically on your finger, any small disturbance could have triggered the fall. Essentially, the crash is the cul-mination of the progressive maturation of the bubble towards its final unstable state. This is often misunderstood and a great controversy starts about its causes.

In order to understand stock markets, one therefore needs to consider the impact of positive feedbacks via possible technical as well as behavioral mechanisms such as imitation and herding, leading to self-organized cooperation and the development of possible endogenous instabilities. The above-sketched universal scenario emphasizing the role of positive feedbacks is at the foundation of our research. Technically, we define a bubble as the "super-exponentially" accelerating rise of a price due to the progressively increasing build-up of cooperation and interactions between investors. The ascent due to positive feedback (also called "pro-cyclicality") is the translation of the maturation towards an instability or critical point reached in finite time. It is a mathematical certainty that, when the growth rate itself grows, the process will be fundamentally unstable. According to this "critical" viewpoint, the specific manner by which the bubble bursts and the prices collapse is secondary: a crash occurs because the market has entered an unstable phase and any small disturbance or process may reveal the existence of the instability. Our research suggests that most of the crashes have fundamentally an endogenous, or internal, origin and that exogenous, or external shocks only serve as triggering factors. As a consequence, the origin of crashes is probably much more subtle than often thought, as it is constructed progressively by the market as a whole, as a self-organizing process. In this sense, the true cause of a crash could be termed a systemic instability. In the sequel, we will review the main bubbles that punctuated the development of the 30 years of the perpetual money machine era, using in each case our models to calibrate financial data, as a way to diagnose the presence of the bubbles. In each of the following figures, the results of our bubble model calculations are shown as lines, ever increasing and faster oscillating towards the critical time, which is the moment of the crash. We stress that these analyses have been performed ex ante (except for the bubble ending in 1987, where the analysis is ex-post), that is, at a time, before the peak and before the subsequent crash confirmed by its occurrence the existence of a bubble.

### 3.2.2 The crash of October 1987: first symptoms

Twenty-five years have passed since the worldwide crash of October 1987, with Black Monday on the 19th that rocked the U.S. markets. For many, its origin is still a mystery, but of little importance, given the prompt recovery that followed. For us, it was the first strong warning signal of the unsustainable growth policies of the 1981–2007 pre-great recession and crisis era (as well as of the new era started in 2008) based on debt and credit that fueled at least seven massive bubbles

(see below), and that led to the current crisis. This is readily seen from the fact that the crash occurred after a period of more than five years of accelerating unsustainable (super-exponential) appreciation, which is characteristic of the first large bubble of that period[5] (see figure 9). It should not be forgotten that this crash rocked global markets, from East Asia to southern America and Europe into the first dramatic worldwide stock market collapse.

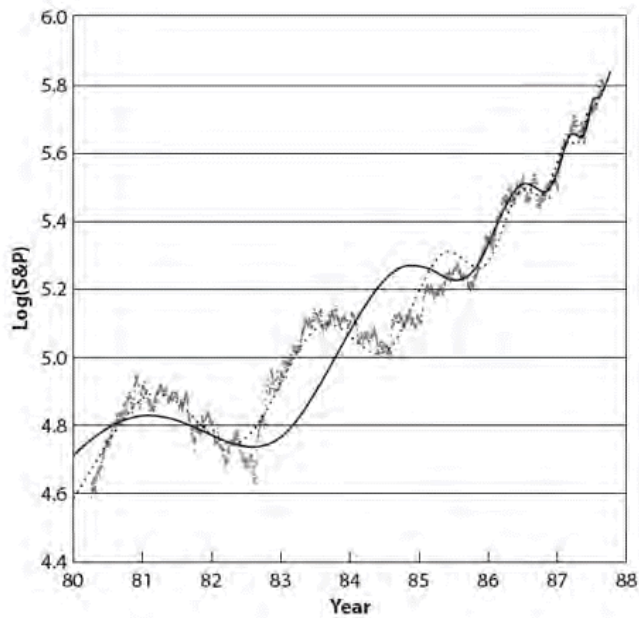

*Figure 9: The historical evolution of the S&P500 index showing the price growing faster than an exponential up to its tipping point of 19th October 1987. The dotted and solid smooth black lines are the results of our model calculations with two different levels of nonlinearity. Reproduced from the book "Why stock markets crash" by D. Sornette[5].*

### 3.2.3 The ICT (dotcom and biotech) "new economy" bubble

The stock markets promptly recovered from the crash of October 1987, particularly in response to the aggressive monetary easing of the Federal Reserve. This led to some turbulence in 1990, but the next massive bubble waited until 1995 to really take off. Over the following five years, the Nasdaq Composite index was multiplied by a factor of five (which corresponds to a 38% average annualized rate of return). In the last year of the run up, it even doubled (that is a 100% annualized rate of return). This acceleration of the rate of return is typical of the super-exponential growth that we characterize as a bubble. In that sense, the dotcom boom was surely one of the best illustrations of the universal pattern behind bubbles and crashes described above. Figure 10 gives the historical evolution of the Nasdaq Composite index in those years.

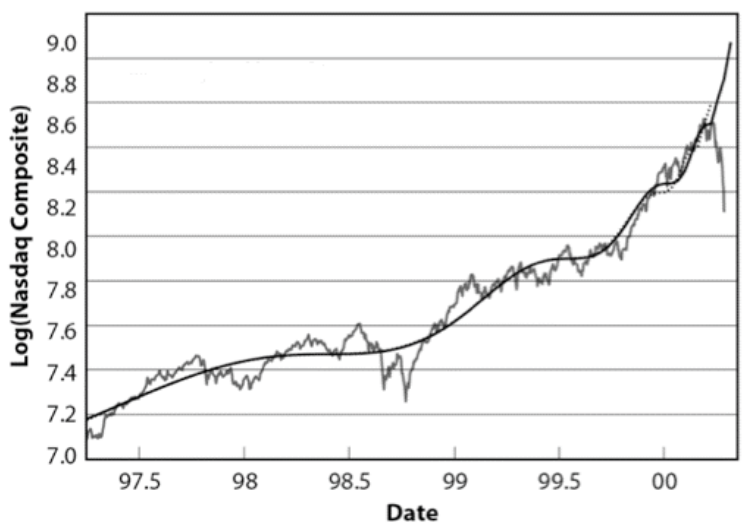

*Figure 10: The historical evolution of the Nasdaq composite index showing the price growing faster than an exponential up to its tipping point and the following crash, starting in March 2000. The solid smooth black line shows the result of our model calculations. Reproduced from the book "Why stock markets crash" by D. Sornette[5].*

*3.2.4 Slaving of the Fed monetary policy to the stock market*

Like in the wake of the 1987 crash, the Federal Reserve re-acted aggressively to stem the recession that followed the bubble's implosion, by decreasing the federal funds rate from 6.5% in 2000 to 1% in 2003 and 2004. Figure 11 shows this process and compares it with the decline of the S&P500 and the Nasdaq Composite stock market indices. It is quite noticeable that the benchmark rate decreased in parallel to the U.S. stock market. The key question, however, is this: did it lead or lag? According to common wisdom and standard textbook arguments, a decrease of interest rates is supposed to make borrowing cheaper. This results in increased expec-tations of future growth. Additionally, lower rates mean lower discount factors. The combined effect should be a boost on stock prices. In this line of reasoning, a federal funds rate decrease should lead economic growth and cause stock market prices to increase. Thus, anti-correlation and a lead of the Fed rate are expected. A quantitative analysis, published by Zhou and Sornette in 2004[6], showed clear evi-dence of exactly the opposite of this textbook argument.

First, a clear correlation is observed. Second, they proved that a lag of about 1 to 2 months existed between the feder-al funds rate and the S&P500 stock market index in which rates were consistently cut after (and not before) stocks de-clined. This study revealed that the Fed monetary policy was influenced significantly by the vagaries of the stock market. This "slaving" of the Fed to the stock markets has been af-firmed in a recent study[7], analysing the central bank's re-sponse to the financial crisis that started in 2007, using a novel methodology allowing for time-varying lead-lag rela-tionships. As a case in point, the rebound in March 2009 of the U.S. stock markets is clearly related to the monetary pol-icy and is especially the consequence of the Quantitative Easing mechanism. In fact, in various writings, speeches and TV interviews, both previous and present Fed chairmen Greenspan and Bernanke have increasingly made clear that the Federal Reserve does care more and more about the evolution of the stock markets and its supposed potentials for wealth creation.

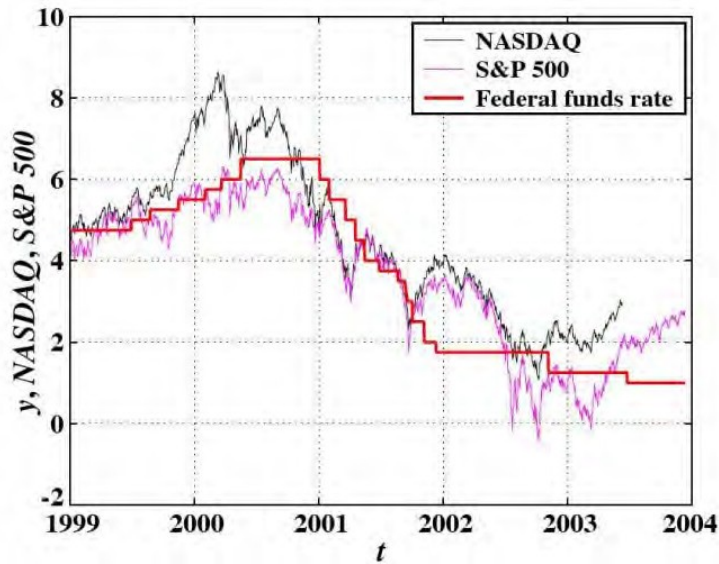

*Figure 11: Comparison of the federal funds rate, the S&P500 and the Nasdaq Composite indices, from 1999 to mid-2003. The time series have been scaled vertically to allow for illustrative visual comparison. Reproduced from Zhou and Sornette (2004)[6].*

*3.2.5 A cascade of bubbles from 2003 to 2008: Real estate, financials, equities, commodities and debt*

This loose monetary policy of the Federal Reserve, together with expansive Congressional real estate initiatives, triggered a chain of events, specifically characterized by a cascade of bubbles.

As can be seen in figure 12, a real estate bubble grew in the U.S. between 2003 and mid-2006. In June 2005, a warning was issued by Zhou and Sornette in a paper distributed on the Cornell University Library website[8] and published later[9]. In the article, they predicted that U.S. house prices would peak mid-2006, basing their analysis on a newly developed bubble diagnostic tool that targets signatures of faster-than-exponential growth.

When real estate prices came down in the U.S. in 2007, the impact was of massive proportions. A structured credit boom had developed in parallel with the house price bubble. As a consequence, lending standards had deteriorated to unseen levels. This is well illustrated by the occurrence of so-called NINJA loans. As a text-book example of what Hyman Minsky called the "Ponzi Borrower", people with No Income, No Job and no Assets were given credit to buy a house. These loans were expected to be paid off by the infinitely rising house prices. Besides, the loans were immediately securitized, and distributed as Mortgage Backed Securities throughout the plumbing of the financial system, eagerly bought by investors. The following implosion of the entangled real estate and structured credit bubbles brought the financial system to the brink of a total collapse.

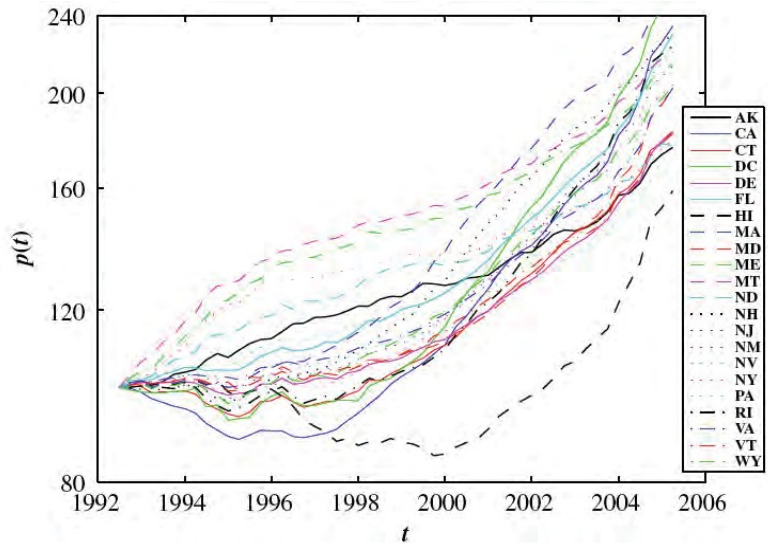

*Figure 12: Quarterly average House Price Index (HPI) in the 21 states and in the District of Columbia (DC) where Zhou and Sornette (2005) diagnosed bubble-type behaviors and predicted a peak in mid-2006. For comparison, the HPI has been normalized to 100 at the second quarter of 1992. The corresponding state symbols are given in the legend on the right[8,9].*

The exuberance, catalyzed by loose monetary policies and the illusion of the perpetual money machine described before, spilled over to all asset classes. This resulted in a stock market, commodities and an energy bubble. Figures 13 and 14 present the historical evolution of the S&P500 index and of the oil price over this period. A clear overall upward curvature can be observed. As we explained before, this is characteristic of an unsustainable super-exponential growth process. Both figures show the calibration of our proprietary models. The good fits, as well as correct early warning signals, confirm the existence of the bubbles. Additionally, an estimated time is given for the expected correction. This is marked by the grey-shaded area in the plots.

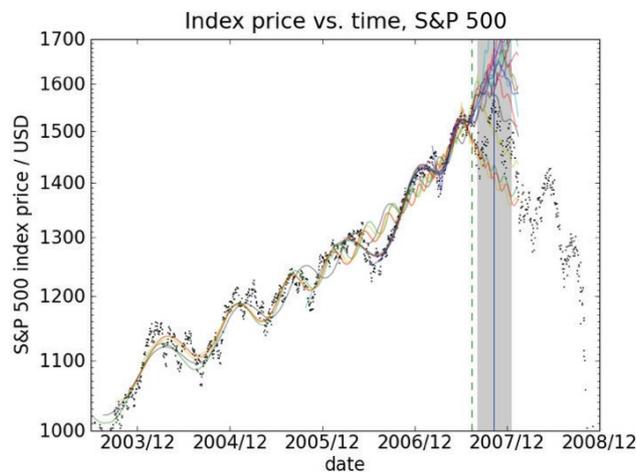

*Figure 13: The historical evolution of the S&P500 index shown as dots. The dashed vertical line shows the last observation used to calibrate our model. The colored curves show different fits. It is expected that the correction occurs with 80% probability in the grey shaded zone. This indeed happened[10].*

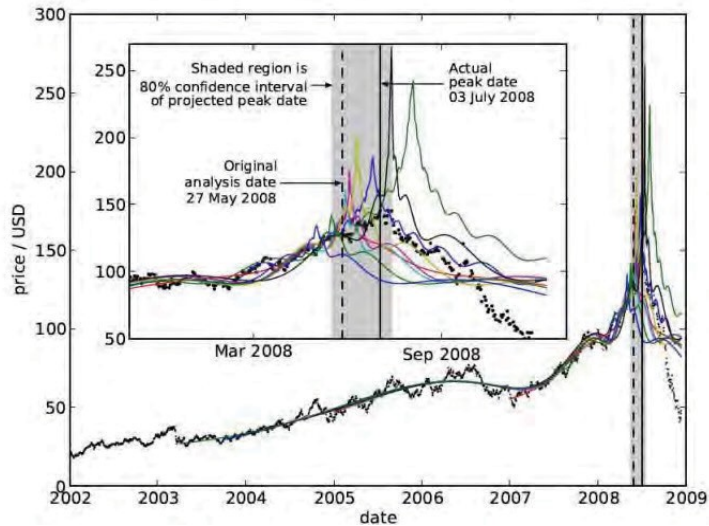

*Figure 14: The historical evolution of oil prices shown as dots. The dashed vertical line shows the last observation used to calibrate our model. The colored curves show different fits. The inset shows a magnification around the time of the peak. As in figure 13, it is expected that the correction occurs with 80% probability in the grey shaded zone. This indeed happened. Reproduced from (Sornette et al., 2009)[10].*

In the end, asset prices worldwide got infected and a globalization bubble emerged, as illustrated in figure 15. The time series represent a proprietary index based on emerging markets equities and currencies, freight prices, soft commodities, base and precious metals and energy.

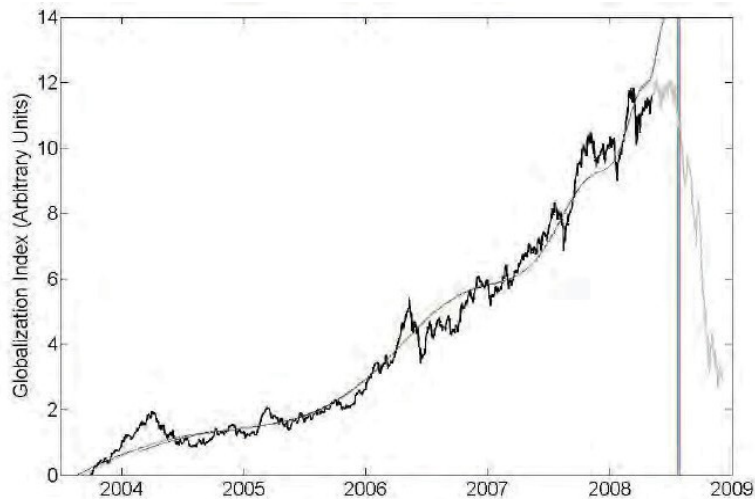

*Figure 15: The globalization bubble. The time series represent a proprietary index of emerging markets equities and currencies, freight prices, soft commodities, base and precious metals and energy. The smooth curves show the fit of the model, the vertical green line is the best estimate of the correction time. Only the black data was used to calibrate the model back in 2008.*

This cascade of bubbles prepared the whole economic and financial systems for a perfect storm. When the final correction came, the global system fell into what is now called the Great Recession, the worst contraction since the Great Depression of the 1930s. This prompted central banks worldwide to stimulate the economy by cutting interest rates and printing money.

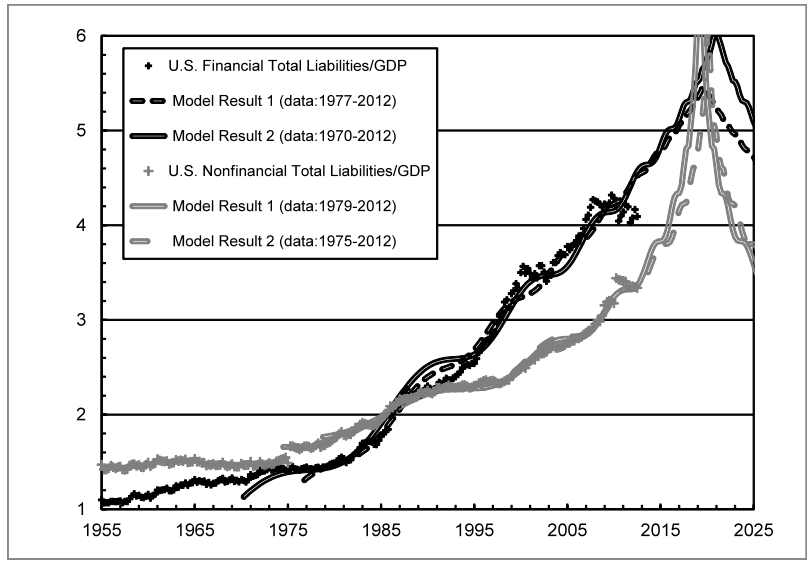

*Figure 16: The total liabilities divided by the GDP of the U.S. financial and non-financial sectors. The data are taken from the Flow of Funds accounts of the U.S. (http://www.federalreserve.gov/releases/z1/), the non-financial sector in- cludes the federal government, government sponsored entities, household and non-profit and non-financial business. The smooth curves show the fits of the models.*

Thirty years ago, our economic and financial system shifted from a growth based on productivity to a growth based on debt. As a consequence, the past three decades have been characterized by financial markets, central banks and treasuries being entangled in a tango of manias and panics. This process has been fueled by ever-increasing debt levels. This is illustrated in figure 16, which shows the evolution of the total U.S. financial and non-financial liabilities divided by the GDP. This picture demonstrates that debt levels are on unsustainable tracks that, according to our bubble models, are expected to reach a critical point towards the end of the present decade.

# 4-The faces of the future

Summarizing, the perpetual money machine is the illusionary belief that it is possible to create wealth out of nothing. At the basis of such wealth creation is debt and the multiplication mechanisms that facilitate credit expansion. The process is self-reinforcing and leads to bubbles. The "machine" was made possible by deregulation of the financial sector and was fueled by quantitative easing policies, which provided easy access to money for financial organizations. Thus, the increased consumption, occurring together with decreasing savings and productivity, becomes a symptom of a growing and dangerous unbalance: an instant purchasing ability was wishfully taken as the confirmation of the existence of an illusory wealth.

The next big question is how long will this perpetual money machine continue operating, and in what form? In this respect, three aspects will clearly differentiate the next decade from the past one: a total financialization of assets in the form of Exchange Traded Funds (ETF), the dominance of the algorithmic trading machines and the future evolution of public debt.

## 4.1 Exchange Traded Funds

ETFs are investment funds traded on stock exchanges. They can replicate the performance of any possible asset like stocks, bonds, commodities or currencies. Before them, there were tracker funds and certificates written on indices. ETFs are attractive because of their low costs, stock-like features and liquidity. More exotic types track specific structured features of assets such as the slope of the yield curve, the implied volatility of a stock market index or even spreads between assets. Many ETFs are leveraged. The daily performance of the ProShares Ultra Silver fund (ticker AGQ) for example corresponds to twice that of silver bullion. In addition, it is perfectly possible to buy options on ETFs. As such, one can buy an option on VXX, which in itself replicates the implied volatility of the S&P500 index options. ETFs represent the ultimate financialization and expectations are high. The size of the ETF market has grown rapidly from around 202 ETFs worth 105 billion USD in 2001 to 4450 ETFs worth 1.5 trillion USD in the first quarter of 2012 (according to BlackRock). Many observers believe this market will grow to 2 trillion USD by 2013, 5 trillion USD by 2015 and 10 trillion USD by 2020.

But what will be the consequences of the growth of ETFs on the structure and stability of our financial system? Many specialists believe that ETFs have in the past years been behind unexpected price moves in physical commodities markets such as copper, cocoa and coffee. There is no doubt about the existence of strongly increased correlations between stocks and general assets in the last few years, making diversification more difficult, if not impossible. Basically, ETFs create new links between different sectors. As such, the constitution of the financial markets is fundamentally changed from an ensemble of individual networks to a closely linked network-of-networks configuration. Very recent research in network theory has shown that, when different networks become linked, the overall structure loses resilience. Indeed, due to diversification, there are less minor events but, due to the stronger coupling, there are more catastrophic events. Additionally, our analyses have shown strong evidence of increased endogenous trading and speculation, which influence soft commodity prices, and which may have a significant impact in poor countries. In this respect, the New England Complex Systems Institute in Boston has published an important analysis[11] showing that the Arab Spring movements started with the second large peak of food prices in 2011 that stressed severely the poor to lower middle classes in the Middle East as well as in many Asian countries. We should always remember that history is replete with political shifts, crises, revolutions and wars caused (or catalyzed) by economic duress (and hunger). The ultimate financialization offered by ETFs opens the road to more bubbles created by the herding mechanism of small investors crowding in and out of the investment fashion of the moment. With the system flushed with liquidity and investors hunting for profit in a zero-interest environment, the frequency of bubbles of all kinds can be expected to increase. Our own analyses at the Financial Crisis Observatory at ETH Zurich have already quantified tens of intermediate sized bubbles in several asset classes in the last few years.

## 4.2 The rise of the algo's

There are lots of talks about the increased importance of the algorithmic trading machines. This new trend started in the U.S. It is now estimated that up to 80% of trading in U.S. stocks is done by machines without any human intervention or decision, of which high frequency trading (HFT) constitutes more than half of the total volume. Meanwhile, algorithmic trading is progressively taking over the global markets, including emerging markets stocks, commodities and energy. The data technology company Nanex has completed an illuminating analysis showing the "rise of the HFT machines" in the past 5 years [12], which warns about the phenomenon of high frequency quoting, namely the fact that 99% of quotes are removed within one second without being executed, in a war game of hide-and-seek between machines. Execution transparency, both in locality (Over-the-Counter (OTC) and dark pools) and with respect to National Best Bid and Offer (NBBO), is gone, which reflects a growing lack of confidence in markets.

Using models specifically designed to quantify the degree of endogeneity (called "reflexivity" by Georges Soros) in the markets, defined as the fraction of transactions that are triggered internally or are self-excited, like aftershocks of an earthquake, and that are not the result of some new external information, we recently quantified that this degree of reflexivity increased from 30% in the 1990s to at least 80% as of today [13]. This proves, in hard numbers, that markets in-creasingly live a life of their own, disconnected from the real economy, activated by machines and the algorithms that compete to trade in milliseconds, a process also facilitated by massive injections of liquidity and the low interest rate policy operating at a different time scale. As a consequence, the bubbles and crashes, that we have become accustomed to, now develop and evolve increasingly over time scales of seconds to minutes. Just look at the flash crash of May 6, 2010, when the Dow Jones Industrial Average fell by 600 points in five minutes. Many other similar sudden drops, followed by fast total or partial recoveries, have occurred since. The logic of endogeneity, reflexivity, positive feedbacks and herding is more and more permeating through all time scales and all instruments in the financial sphere. We do not expect that the technological race will provide a stabilization effect, overall. This is mainly due to the crowding of adaptive strategies used by algorithmic agents, which exhibit pro-cyclical properties (for instance via a preference in so-called momentum trading) and a propensity to herd that is even larger than found in human beings. No level of technology can change this basic fact, which is widely documented for instance in artificial worlds populated by software-agents that simulate financial markets on computers. New algorithms that exploit the high volatility periods associated with distress and crashes are been vigorously developed. These are really worrying trends.

## 4.3 The future of public debt

We made clear above that economic growth in the last 30 years has been essentially fueled by explosive debt growth. The future of debt cannot be assessed correctly without the perspective provided by the history of its first 5000 years, reconstructed by David Graeber in his recent book [14]. Graeber sees the capitalistic age (1450–1971) as a return to the quantification and systems of slavery and debt peonage. According to him, the new era that started in 1971 is characterized by a resistance toward debt peonage and the emergence of the dominance, not of monarchs, but of financiers and banks. The fundamental insight is the fact that banking has taken over the role of monarchs in the system of debt peonage, even binding governments in debt. Importantly, Graeber shows that debt has been the unifying underlying process by which groups and societies have developed, strived and collapsed. Human debt history is cyclic, with periods of unbridled debt growth, which seem intrinsically unstable, followed by biblical like jubilees. The present travail in Europe in particular seems to be balancing between disguised steps towards debt pardon and attempts to prolong the status quo.

Therefore, in addition to the expected impact of financialization and automatization in changing the face of our financial markets and consequently of our economy in the decades to come, an even larger revolutionary change can be expected from the future evolution of government debts. In a report, published by the Bank of International Settlements in 2010[15], Stephen G. Cecchetti, M.S. Mohanty and Fabrizio Zampolli, go through great lengths to explain the inherent risk in the future of the public debt trajectory. Figure 17 summarizes their Debt/GDP projections, over the next 30 years, for a dozen major industrial economies. It is clear that, if no substantial measures are taken, the debt burden of the public sector will grow to unsustainable proportions. Just keeping the status quo in the face of existing excesses and liabilities requires that debts grow even faster. One additional aggravating actor that is going to play in the coming decades, on top of the continuation of the perpetual money machine thinking, is the future costs, like pensions and medical aid programs, coming from the rapidly ageing population. There are no savings put aside to finance these costs. That is why they are often referred to as unfunded liabilities. It is clear that they will have to be managed by a long-term fiscal planning, which will inevitably come with a reassessment of our expectations.

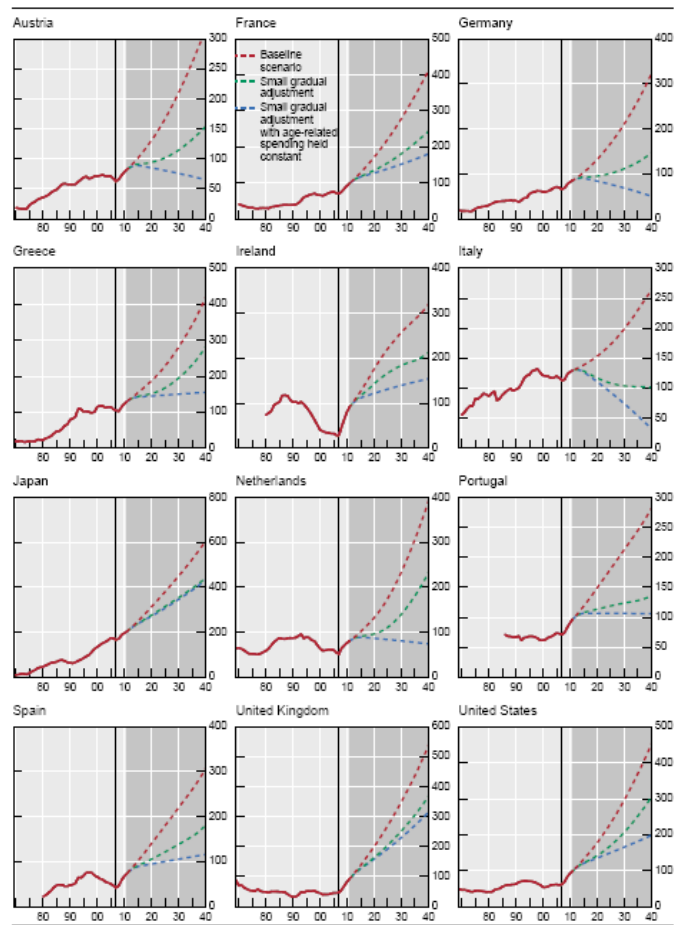

Sources: OECD; authors' projections.

*Figure 17: The projection of Debt/GDP in a dozen of major industrial economies following three different scenarios, source: BIS Working Papers No 300[13].*

### 4.4 Final diagnostics

Like the Red Queen of Lewis Carroll's novel "Through the Looking-Glass" telling Alice that "It takes all the running you can do, to keep in the same place", central bank interventions seem to just achieve that, the status quo. Therefore, the real question is whether all these activities provide genuine value. Our research and that of many others do not support the Wall Street view that more financial transactions benefit the real economy. What is becoming very clear, however, is that the deregulation of finance and banking has created much more instability. It has addicted the real economy to financial steroids. Banks have no incentive to align the growth of credit and the volume of financial transactions with the real economy. This conflict of interests will further prolong and enhance our problems. The Glass-Steagall Act of 1933 addressed the destabilization nature of the financial system by separating investment, commercial and retail banking and insurance and by designing incentives favoring the real economy. Many scholars attribute much of the financial stability after the Second World War to this legislation. Its repeal in 1999 was the culmination of a decade of deregulation justified by the 'great moderation', which turned out to be just a consequence of the actuation of a perpetual money machine dream that promised unrealistic economic growth.

Extrapolating the evidence gathered until now, absent fundamental collective actions to stop the toxic logic of the perpetual money machine, a rather gloomy future is coming into focus, with more bubbles and crashes occurring at all-time scales, increasing systemic instability, dilution of wealth via disguised inflation, unsustainable unreal economic growth and exploding debts, coming with growing inequalities and social unrests. The science of instabilities has progressed enough that one can shape clear diagnostics. The good news is that solutions are known and are a matter of political will and social contract. We need to go back to a financial system and debt levels that are in balance with the real economy. Rather than still hoping that real wealth will come out of money creation, an illusion also found in the current management of the on-going European sovereign and banking crises, we need a fundamentally different approach. Evidently, we need real, non-financial economic growth based on improved infrastructure, innovation, technology, creativity, all fueled by vigorous investment in human capital, research and development. Easier said than done in a world whose ability to innovate and to grow is now stalling due to over-capacity, over-indebtedness, an oversized and monopolistic banking system, an obsolete taxation system, socially perverse incentives and the influence of lobbies paralyzing necessary change.

# 5-What can we do?

Given the above analyses and diagnostics, it is all too tempting to propose sweeping changes. In this final part, we stress how dangerous are sudden changes in the management of a complex system. The difference between the firefighting strategies in the U.S. and Mexico makes a beautiful case-in-point that we will explain later. Only gradual change, with a clear long term planning, can steer our financial and economic system from the turbulence associated with the perpetual money machine to calmer and more sustainable waters. As a consequence, there will be continuing uncertainty in the years to come. Investors should focus on real value as well as the recognition of burgeoning bubbles for the preservation of their wealth. Governing is the art of planning and prediction. In uncertain times, it is essential, more than ever, to think in scenarios: what can happen in the future, and, what would be the effect on your wealth and capital? How can you protect yourself and your dearest against adverse scenarios? In the following, we examine the question "what can we do?" from the macro level, discussing the fundamental issue of incentives and of constructing and predicting scenarios. We then conclude with investment insights.

## 5.1 It is all about incentives

All discussions about change are superficial and naïve if they do not touch upon the fundamental drivers of people, their incentives. Unsustainable situations are often caused by misaligned incentives. There are many examples, from the loss of the fiduciary principle [16], and the reward of financial managers in relative rather than absolute value terms enhancing the short term impact of luck and randomness at the expense of long-term economically meaningful investments, the encouragement of corporate management to misreport and develop frauds by the existing legal system allowing unbridled retroactive lawsuits (of course, accountability and responsibility for errors from the past is essential but not as done today), all the way to the medical and pharmaceutical industries that have the rational economic incentive to keep us all marginally ill [17]. To get out of this catch 22 situation, a major overhaul of the incentive systems of our societies should be the priority. Guidance may be obtained for instance from Memorial Nobel prize winner (2009) Elinor Ostrom's work on collective action to ensure long-term sustainable resource yields based on design principles promoting stable bottom-up management [18]. Examples of concrete applications where the collective action approach has shown positive results include the management of resources extracted from forests, fisheries, oil fields, grazing lands, and irrigation systems in many parts of the World.

## 5.2 Managing the transition by planning and predicting

Richard Werner, of the University of Southampton in the UK, presents in a recent paper [19] a nice example of how better monitoring and management systems could substantially improve the resilience of our economic and financial system. In his research, Werner makes a clear distinction between credit that boosts the real economy and credit that fuels asset prices and bubbles. Banks are the creators of money through credit. Werner proposes that this process should be carefully monitored, so that we have a better understanding of which money is used for the real and which is used for the financial economy. He takes this even one step further. A system of what he calls "credit guidance" should be designed by the central banks. This could steer credit and newly created money in the direction of the real economy, to be used to increase productivity and GDP-growth. In Germany for example, this "credit guidance" comes naturally from the structure of the banking system itself. There, 70% of the deposits are accounted for by over 1,000 locally headquartered small savings and coopera-tive banks. These institutions focus on lending to the household and productive SME sector. While less technically sophisticated than the international banks, these small banks show core competence by knowing their customers. A better monitoring and management of the credit and money creation process, together with a banking sector that is naturally interwoven in the real economy, will certainly make our economic and financial system much more resilient.

This common sense approach is part of the solutions to our problems, by putting emphasis on the simple rather than on the complex. But the present trend towards more and more complexity, as witnessed in the ever more complex regulations, is not a good sign. Take the Dodd-Frank act of 2010 of financial regulations, which runs to 848 pages (compared with 30 pages for the Glass-Steagall act of 1933), with almost 400 more pieces of detailed rule-making for implementation and less than half of the rules being finalized at the time of writing (October 2012).

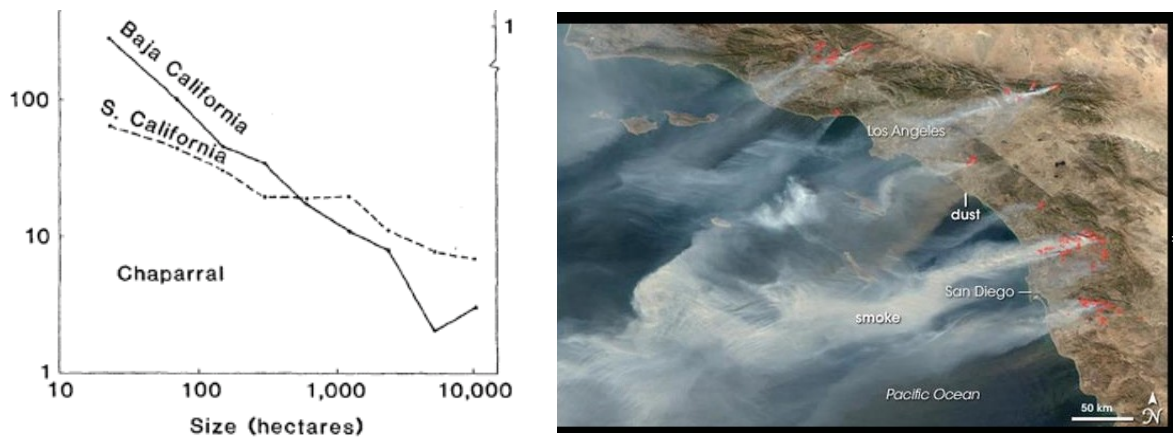

*Figure 18: Left: Number of fires plotted against burn area for Baja Califoria (Mex- ico) and Southern California (U.S.) from 1972 to 1980. Chaparral is a shrub land or heathland plant community found primarily in the U.S. state of California and in the northern portion of the Baja California peninsula, Mexico. Reproduced from Minnich, 1983 [20]. Right: Space photograph of the October 2007 California wildfires that led to 2000 km2 of land burned from Santa Barbara County to the US-Mexico border.*

The impact of simple versus complex methods is perhaps nowhere better illustrated than in the distinct management styles of forest fires in Southern California in the U.S. compared with Baja California in the North-Western part of Mexico, two regions with essentially the same climate and vegetation. In the former, the objective is to stop all fires. In the latter, it is laissez-faire, arguably due to weaker resources and smaller loss exposure. The results are drastically different as shown in the left panel of figure 18: while Southern California has few small fires, extremely large fires are shockingly frequent, corresponding to conflagrations that occur when a set of negative factors combines together (e.g. the accumulation of wood fuel, hot weather, strong wind) that no human efforts can stem as exemplified by the October 2007 California wildfires; in contrast, Baja California is graced with many small and essentially no large fires. Inspired by this observation, the laissez-faire strategy was applied to fire management in Yellowstone Park. But to the dismay of the decision makers this turned into an inferno. What was forgotten is that the favorable situation in Baja California stems from the self-organized dynamics between forest growth and fires that keep them in balance and that has evolved over many decades. In contrast, the laissez-faire strategy was transplanted to the Yellowstone Park after decades of strong fire reduction efforts, placing the system in an unstable state with large accumulated wood fuel. As a consequence, the initial conditions were not adapted and the strategy backfired (pun intended). By this example, we learn that one cannot change a system without risking cat-astrophic readjustments. The financial and economic systems are addicted to financial steroids, and we need a gradual and careful approach to manage the change from the perpetual machine debt approach to a more sustainable regime.

"Gouverner, c'est prévoir" (Governing is the art of planning and predicting). This quip by Émile de Girardin, the 19th century French journalist and politician, summarizes a possible approach for steering this transition, namely through the development of Observatories for economics and finance, built on an interdisciplinary consortium of economists, natural scientists, computer scientists and engineers, to combine their expertise to forecast and anticipate looming instabilities. Only by measuring, monitoring and predicting can one really shape an informed view to manage and govern. But, unfortunately, economics and finance have been stalled by a general dogmatic refusal to embrace the technology of the natural sciences, including its rigorous validation process. As a case in point, it is shocking that essentially none of the economic and financial forecasts routinely provided by the many public and private agencies in all countries, which are used by decision makers and government to shape their policies, are examined ex-post to assess their predictive merits. We are like the elegant butterflies flying in the dark, attracted to the illusory light of prediction, burning ourselves with repeated errors, just to fly again and again toward the illusory light. Never learning, always active, finding solace in our hyper-activity, however poor its track record, which is anyway ignored in the flurry of the present. The Financial Crisis Observatory at ETH Zurich launched in 2008 has the clear objective of filling up this gap, by monitoring and diagnosing financial bubbles and instabilities. Even more ambitious is the European Economic and Financial Exploratory that is contemplated within the FuturICT large scale European research initiative [21] which aims at continuously monitoring and evaluating the status of the economies of countries in their various components, developing new reliable diagnostic tools and providing the framework to perform what-if analyses and scenario evaluations to inform decision makers and help develop innovative policy, market and regulation designs.

## 5.3 Investment insights

What about investments? How can a responsible investor protect herself against the possible hardships ahead? In our view, this should be done by planning and predicting; thinking in scenarios and focusing on protection and preservation.

A first essential ingredient is the need for what we call a "time-at-risk" approach. This takes into account the intrinsic non-stationary nature of financial markets and the importance of monitoring bubbles to diagnose the critical points associated with their bursts. This can lead to investment strategies that profit from the knowledge of bubbles.

Secondly, in a context of zero short-term interest rates, almost unprecedented low long-term yields and the specter of inflation, the global insight and landscape painted above suggest a focus on real value, ensuring long-term sustainability, i.e. the protection of capital rather than the quest for high returns. Focusing on physical commodities and what is really needed in the economy on the medium and long term (industrial metals, energy, food) or natural resources (like the Harvard endowment trust investing in Romanian Forests) seems absolutely necessary.

In a world of galloping sovereign debts, some corporate debts clearly provide safer investment than government bonds. Additionally, with the (non-official but real) focus of central banks on stock markets, it would be unreasonable to stay out of them for their potential for appreciation, to stay in place in real purchasing power.

Small and medium size firms with value (large book-to-market) have historically delivered the highest returns, even when adjusted for survival biases. There is no better time than in periods of uncertainties to be "contrarian" and invest in entrepreneurial projects, with a rigorous diligence applied to the detection of innovations and the talent and passion of the entrepreneurs and directors. Many of the great companies, such as Apple, General Electric, IBM, Hewlett-Packard, and Microsoft were launched at times of recessions.

A delicate balance on a case-to-case basis has to be found between investment in (private) equity to provide access to real company investments and corporate credit to give "credit to the real economy". The choice of duration should not be "one clothe for all" but chiseled to each specific bond structure.

There is also the possibility to be "long" core inflation (CPI) in the form of TIPS (Inflation protection Treasuries). However, the large demand for this type of instruments makes them more and more de-correlated to inflation, a signature of an arbitrage opportunity that is disappearing. In the same spirit, due to the concerted action of governments and central banks, financial volatility seems quite cheap and can provide another edge against the future turbulences.

In conclusion, our main message is that our economy has become ever more financialized in the past three decades in an unsustainable illusion of a perpetual money machine. The corollary is that true and sustainable value is more likely to be found in sectors that have the potential for real growth of production through new technologies, innovations and creativity such as biotechnology and health care.

## Endnotes

1  *Alan M. Taylor, The Great Leveraging, University of Virginia, NBER and CEPR, July 2012*

2  *Debt and deleveraging: Uneven progress on the path to growth, McKinsey Global Institute, January 2012*

3  *Trends.google.com*

4  *Charles Kindleberger, Manias, Panics and Crashes, Wiley (1978)*

5  *Didier Sornette, Why Stock Markets Crash Critical Events in Complex Financial Systems, Princeton University Press, January 2003.*

6  *W.-X. Zhou and D. Sornette, Causal Slaving of the U.S. Treasury Bond Yield Antibubble by the Stock Market Antibubble of August 2000, Phys- ica A 337, 586–608 (2004)*

7  *Kun Guo, Wei-Xing Zhou, Si-Wei Cheng and Didier Sornette, The U.S. stock market leads the Federal funds rate and Treasury bond yields, PLoS ONE 6 (8), e22794 (2011)*
   *(http://dx.doi.org/10.1371/journal. pone.0022794)*

8  *http://arxiv.org/abs/physics/0506027*

9  *W.-X. Zhou and D. Sornette, Is There a Real Estate Bubble in the US? Physica A 361, 297–308 (2006)*

10 *D. Sornette, R. Woodard and W.-X. Zhou, The 2006-2008 Oil Bubble: evidence of speculation, and prediction, Physica A 388, 1571–1576 (2009)*

11 *Marco Lagi, Karla Z. Bertrand, Yaneer Bar-Yam, The Food Crises and Political Instability in North Africa and the Middle East, http://arxiv.org/ abs/1108.2455 (Submitted on 11 Aug 2011)*

12 *http://www.nanex.net/aqck/2804.HTML*

13 *V. Filimonov and D. Sornette, Quantifying reflexivity in financial markets: towards a prediction of flash crashes, Phys. Rev. E 85 (5): 056108 (2012)*

14 *Graeber, D., Debt: The First 5,000 Years, Melville House, First Edition edi- tion (July 12, 2011).*

15 *Stephen G. Cecchetti, M.S. Mohanty and Fabrizio Zampolli, " The Future of Public Debt: Prospects and Implications", BIS Working Papers No 300, March 2010 (http://www.bis.org/publ/work300.pdf)*

16 *John C. Bogle, The fiduciary principle (no man can serve two masters), The Journal of Portfolio Management 36 (1), 15–25 (2009)*

17 *In ancient China and in traditional India, the doctor was not paid when one fell sick, only on the days when one was healthy. The role of the doctor was to aid in the maintenance of health. It was considered inhu- man and devilish to take fees from a sick person who was suffering. A doctor was supposed to benefit from a person's health and not from his disease.*

18 *Elinor Ostrom, Beyond Markets and States: Polycentric Governance of Complex Economic Systems, American Economic Review 100 (3), 641-672, June2010.*

19 *Towards a New Research Programme on „Banking and the Economy" – Implications of a Quantity-Equation Model for the Prevention and Resolution of Banking and Debt Crises, Richard A. Werner, Accepted for publication in the International Review of Financial Analysis, Accepted date 9 June 2012*

20 *Richard A. Minnich, Fire mosaics in Southern California and Northern Baja California, Science 219 (4590), 1287–1294 (1983).*

21 *www.futurict.eu*

*Notenstein Academy*

**The Notenstein Academy White Paper Series is published by the Notenstein Academy, the educational and academic platform of Notenstein Private Bank Ltd. In addition to providing in-house training, the Academy serves to promote dialogue on current issues in the financial sector and provide a foundation for grounded, forwards-looking wealth management that is based on thinking in scenarios.**

*Legal notice*